\documentclass[preprint,12pt]{elsarticle}
\usepackage{epsfig}
\usepackage{amssymb}
\journal{NIM Section A}
\usepackage{subfigure}
\usepackage{lineno}
\usepackage{float}
\usepackage{bm}
\usepackage{titlesec}
\titleformat{\paragraph}[runin]
{\bfseries\scshape}{\theparagraph}{1em}{}
\setcitestyle{square}

\begin{document}
\begin{frontmatter}
\date{\today}

\title{Dynamic tunable notch filters for the Antarctic Impulsive Transient Antenna (ANITA)}

\author[OSU] {P.~Allison}
\author[OSU] {O.~Banerjee\corref{cor1}}
\ead{oindreeb@gmail.com}
\cortext[cor1]{Corresponding author}
\author[OSU] {J.~J.~Beatty}
\author[OSU] {A.~Connolly}
\author[Chicago] {C.~Deaconu}
\author[OSU] {J.~Gordon}
\author[UH] {P.~W.~Gorham}
\author[OSU] {M.~Kovacevich}
\author[UH] {C.~Miki}
\author[Chicago] {E.~Oberla}
\author[UH] {J.~Roberts}
\author[UH] {B.~Rotter}
\author[OSU] {S.~Stafford}
\author[UH] {K.~Tatem}

\author[UCL] {L.~Batten}
\author[JPL] {K.~Belov}
\author[KU, MEPHI] {D.~Z.~Besson}
\author[WashU] {W.~R.~Binns}
\author[WashU] {V.~Bugaev}
\author[UD] {P.~Cao}
\author[NTU]{C.~Chen}
\author[NTU] {P.~Chen}
\author[NTU]{Y.~Chen}
\author[UD] {J.~M.~Clem}
\author[UCL] {L.~Cremonesi}
\author[OSU] {B.~Dailey}
\author[UCLA] {P.~F.~Dowkontt}
\author[NTU]{S.Hsu}
\author[NTU]{J.~Huang}
\author[OSU] {R.~Hupe}
\author[WashU] {M.~H.~Israel}
\author[UH] {J.~Kowalski}
\author[UCLA] {J.~Lam}
\author[UH] {J.~G.~Learned}
\author[JPL] {K.~M.~Liewer}
\author[NTU] {T.~C.~Liu}
\author[Chicago] {A.~Ludwig}
\author[UH] {S.~Matsuno}
\author[UD] {K.~Mulrey}
\author[NTU] {J.~Nam}
\author[UCL] {R.~J.~Nichol}
\author[KU, MEPHI] {A.~Novikov}
\author[KU] {S.~Prohira}
\author[WashU] {B.~F.~Rauch}
\author[NTU]{J.~Ripa}
\author[JPL] {A.~Romero-Wolf}
\author[UH] {J.~Russell}
\author[UCLA] {D.~Saltzberg}
\author[UD] {D.~Seckel}
\author[NTU]{J.~Shiao}
\author[KU] {J.~Stockham}
\author[KU] {M.~Stockham}
\author[UCLA] {B.~Strutt}
\author[UH] {G.~S.~Varner}
\author[Chicago] {A.~G.~Vieregg}
\author[NTU]{S.~Wang}
\author[CalPoly] {S.~A.~Wissel}
\author[UCLA] {F.~Wu}
\author[KU] {R.~Young}

\address[OSU] {Dept. of Physics, The Ohio State Univ., Columbus, OH 43210; Center for Cosmology and AstroParticle Physics.}
\address[UCL] {Dept. of Physics and Astronomy, University College London, London, United Kingdom.}
\address[JPL] {Jet Propulsion Laboratory, Pasadena, CA 91109.}
\address[KU] {Dept. of Physics and Astronomy, Univ. of Kansas, Lawrence, KS 66045.}
\address[WashU] {Dept. of Physics, Washington Univ. in St. Louis, MO 63130.}
\address[UD] {Dept. of Physics, Univ. of Delaware, Newark, DE 19716.}
\address[Chicago] {Dept. of Physics, Enrico Fermi Institute, Kavli Institute for Cosmological Physics, Univ. of Chicago , Chicago IL 60637.}
\address[UH] {Dept. of Physics and Astronomy, Univ. of Hawaii, Manoa, HI 96822.}
\address[NTU] {Dept. of Physics, Grad. Inst. of Astrophys., Leung Center for Cosmology and Particle Astrophysics, National Taiwan University, Taipei, Taiwan.}
\address[UCLA] {Dept. of Physics and Astronomy, Univ. of California, Los Angeles, Los Angeles, CA 90095.}
\address[MEPHI] {National Research Nuclear University, Moscow Engineering Physics Institute, 31 Kashirskoye Highway, Russia 115409}
\address[CalPoly] {Dept. of Physics, California Polytechnic State Univ., San Luis Obispo, CA 93407.}

\begin{abstract}

The Antarctic Impulsive Transient Antenna (ANITA) is a NASA long-duration balloon experiment 
with the primary goal of detecting ultra-high-energy ($>10^{18}\,\mbox{eV}$) neutrinos via the Askaryan Effect. 
The fourth ANITA mission, ANITA-IV, recently flew from Dec~2 to Dec~29, 2016. 
For the first time, 
the Tunable Universal Filter Frontend (TUFF) boards were deployed
for mitigation of narrow-band, anthropogenic noise with tunable, switchable notch filters. 
The TUFF boards also performed second-stage amplification by approximately
45~dB to boost the $\sim\,\mu\mbox{V-level}$ radio frequency (RF) signals to $\sim$ mV-level for digitization, and 
supplied power via bias tees to the first-stage, antenna-mounted amplifiers. 
The other major change in signal processing in ANITA-IV is the resurrection of the
$90^{\circ}$ hybrids deployed previously in ANITA-I, in the trigger system, although in this paper we focus on the TUFF boards.
During the ANITA-IV mission, the TUFF boards were successfully operated throughout the flight. 
They contributed to 
a factor of 2.8
higher total instrument livetime on average in ANITA-IV compared to ANITA-III
due to reduction of narrow-band, anthropogenic noise before a trigger decision is made. 

\end{abstract}


\end{frontmatter}

\section{Introduction}

The Antarctic Impulsive Transient Antenna (ANITA) is a NASA long-duration balloon-borne ultra-high-energy (UHE) neutrino detector \cite{instrPaper}.
ANITA looks for radio impulses produced via the Askaryan Effect by UHE neutrinos interacting in the Antarctic ice.
The Askaryan Effect, as formulated by Askaryan \textit{et al}. \cite{askaryan} and observed in ice by the ANITA collaboration in a beam test \cite{askaryan_observation}, is the production of coherent Cherenkov radio impulses due to a 
charged particle shower traveling in a dielectric medium at a speed faster than the speed of light in that medium.

The fourth ANITA flight, ANITA-IV, was launched on Dec~2, 2016 from the NASA Long Duration Balloon
(LDB) Facility located $10\,\mbox{km}$ from McMurdo Station in Antarctica. 
The flight was terminated on Dec~29, 2016 and landed approximately $100\,\mbox{km}$ from the South Pole. 
The Tunable Universal Filter Frontend (TUFF) boards were deployed for the first time in the ANITA-IV mission,
and 
are the subject of this paper. 

\subsection{Continuous-wave (CW) interference} 
The principal challenge of the ANITA experiment is to distinguish neutrino signals from radio frequency (RF) noise. 
The two main sources of noise are thermal radiation by the Antarctic ice and anthropogenic noise, much of which is continuous-wave (CW) interference. 

While Antarctica itself is relatively free of CW transmissions, except for bases of human activity, transmissions from geosynchronous satellites are continuously in view.
The average full-width-at-half-maximum (FWHM) beamwidth of the ANITA antennas is approximately $45^{\circ}$.
Although the ANITA antennas are canted downward by $10^{\circ}$, the beam of the antennas extends to horizontal from the perspective of the payload and
above.  
The Antarctic science bases, the most prominent being McMurdo and South Pole Station, are more radio-loud than the rest of the continent, producing CW interference, for example, in the $430-460\,\mbox{MHz}$ band. 

CW interference due to military satellites has affected all ANITA flights.
ANITA-I (Dec. 2006 - Jan. 2007) and ANITA-II (Dec. 2008 - Jan. 2009) observed CW interference primarily in the $240-270\,\mbox{MHz}$ band, peaking
at $260\,\mbox{MHz}$. This frequency range is predominantly used by the aging Fleet Satellite (FLTSAT) Communications System 
and the Ultra High Frequency Follow-On (UFO) System, both serving the 
United States Department of Defense since year 1978 and 1993 respectively.
In addition to CW interference at $260\,\mbox{MHz}$, ANITA-III (Dec. 2014 - Jan. 2015) observed CW interference at $375\,\mbox{MHz}$
which is thought to be due to
the newer Mobile User Objective System (MUOS) satellites that were launched during the period from Feb. 2012 - June 2016 \cite{milsat}.

The ANITA-III experiment was most affected by CW interference due to military satellites.
Due to the design
of the ANITA-I and ANITA-II trigger, which required coincidences among different frequency bands, the CW interference did not overwhelm
the acquisition system. 
However, ANITA-III was redesigned for improved sensitivity and based its trigger decisions on full-bandwidth ($200 - 1200\,\mbox{MHz}$) signals. 
This produced trigger rates far in excess of the digitization system's readout
capabilities ($\sim50\,\mbox{Hz}$) for thresholds comparable to those used in previous flights. 
Thus, the ANITA-III experiment was susceptible to digitization deadtime (defined in Section~\ref{livetime_section}) throughout the flight. 

The lesson learned from the ANITA-III flight was
that a new method of mitigation of CW signal had to be a priority for the ANITA-IV flight. 
Before ANITA-IV, the available methods to reduce digitization deadtime were
masking
and decreasing thresholds (described in Sections~\ref{processing} and \ref{methods}) when in the presence of higher levels of noise. 
A decrease in thresholds corresponds to higher power of the incoming signal as explained in Section~\ref{trigger}. 
Both of these methods come at a high price.
During the majority of the ANITA-III flight, masking was used during noisy periods to veto triggers from
over 
half of the payload field-of-view to keep the
trigger rate at or below $50\,\mbox{Hz}$. 
This significantly lowered the total instrument livetime (defined in Section~\ref{livetime_section}). 
Decreasing thresholds led to reduction of sensitivity to neutrinos during noisy periods. 
For ANITA-IV, the TUFF boards were built with
tunable notch filters to restore triggering efficiencies 
in the presence of CW interference. 
Additionally, the $90^{\circ}$ hybrids, previously deployed
 in ANITA-I as described in our design paper \cite{instrPaper}, were added to the ANITA-IV trigger system
 to require signals to be linearly polarized. 

\section{ANITA Payload}
\label{payload} 

\begin{figure}[H]
\centering
\includegraphics[width=1.0\textwidth]{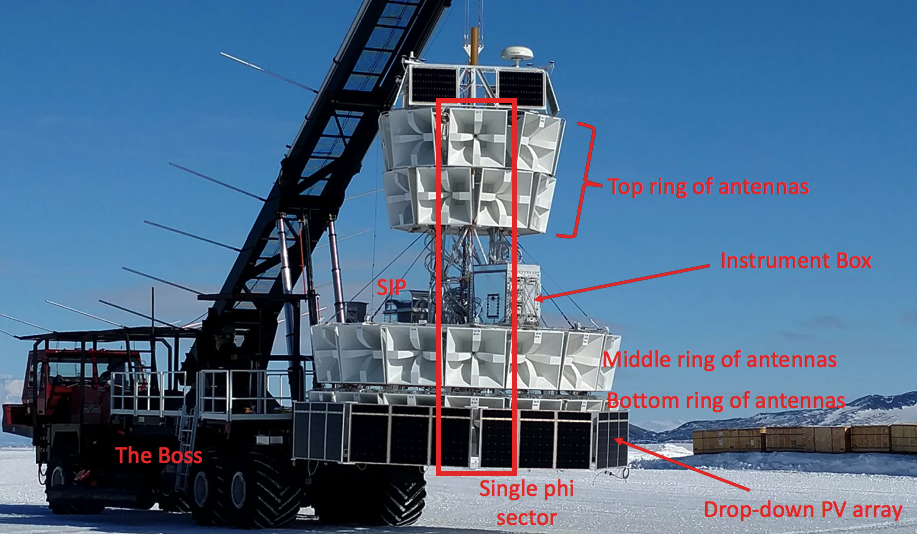}
\caption{The ANITA-IV payload just prior to launch at the NASA LDB Facility near McMurdo Station, Antarctica. The red box encloses three antennas that make up a single phi sector.}
\label{anita}
\end{figure}

The ANITA payload is designed to view the ice out to the horizon at $700\,\mathrm{km}$ distance with complete azimuthal coverage and good reconstruction capability, while its 
shape and size is constrained by its NASA launch vehicle ``The Boss," pictured in Figure~\ref{anita}. 
The ANITA-III and ANITA-IV payloads each have 48 antennas. 
The antennas are arranged in three aligned rings of 16 antennas, termed the top, middle, and bottom rings. 
The top ring consists of two staggered sub-rings each having eight antennas. 

The three rings of antennas and a phi sector of ANITA-IV are pictured in Figure~\ref{anita}. 
The FWHM beamwidth of the antennas is approximately $45^{\circ}$. 
The antennas in the top ring are evenly spaced by $45^{\circ}$ in azimuth. 
The two sub-rings in the top ring are offset by $22.5^{\circ}$ for uniform coverage.
The antennas in the middle ring are evenly spaced by $22.5^{\circ}$.
The antennas in the bottom ring are evenly spaced by $22.5^{\circ}$.
All the antennas are angled downward by $10^{\circ}$ to preferentially observe signals coming from the ice as opposed to from the sky. 
Each group of three antennas in a vertical column, taking one antenna from each ring, forms a phi sector, viewing a $22.5^{\circ}$ region in azimuth. 

\begin{figure}[H]
\centering
\includegraphics[width=1.0\textwidth]{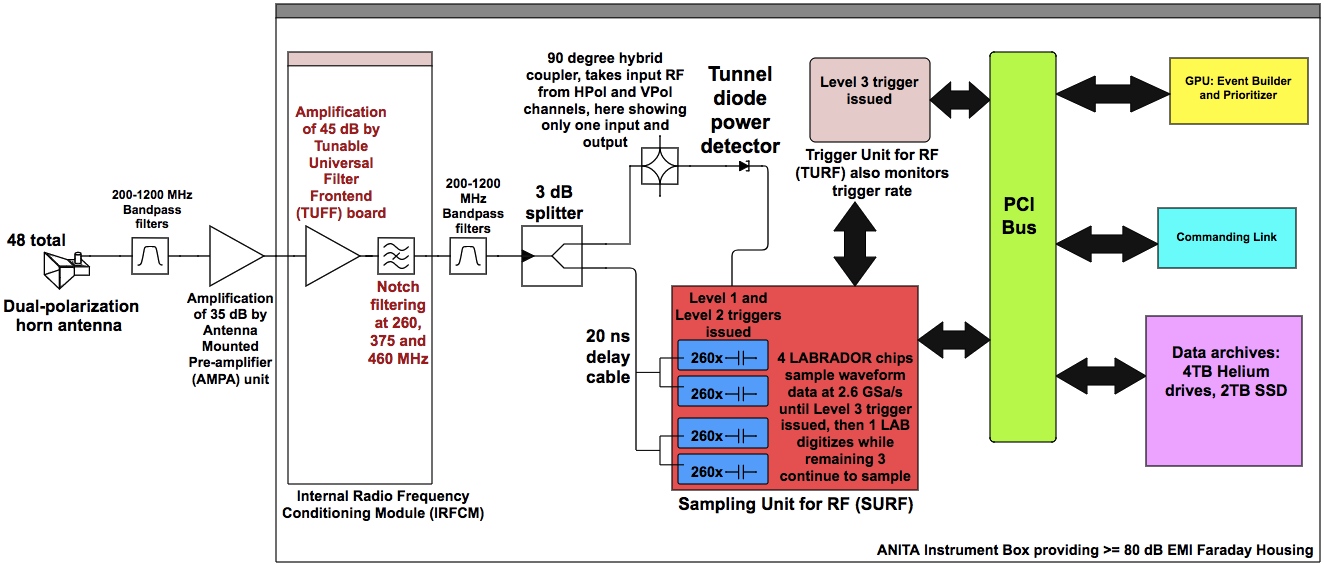}
\caption{The ANITA-IV signal processing chain for a single RF channel.}
\label{system}
\end{figure}

The ANITA Instrument Box is placed on a deck above the middle ring of antennas, also seen in Figure~\ref{anita}. 
The Instrument Box contains different units for signal processing, as illustrated in Figure~\ref{system}. 
More details on signal processing are in Section~\ref{processing}. 
In ANITA-IV, the 12-channel TUFF modules reside inside four Internal Radio Frequency Conditioning Modules (IRFCMs) inside the Instrument Box.

The NASA Science Instrument Package (SIP) also sits on the deck.
The SIP is powered and controlled by NASA.
It is used for flight control such as ballast release and flight termination. 
The SIP also provides a connection to the ANITA payload during flight through line-of-sight transmission, the Iridium satellites, and the Tracking and Data Satellite System (TDRSS). 
This allows us to monitor the payload continuously during the flight.
A small fraction of data (less than 1\%) is transferred from the payload through telemetry.
Commands to perform different functions, such as tuning a TUFF notch filter, 
can be sent to the payload in real time using the SIP connection. 

\section{ANITA Signal Processing} 
\label{processing}

In this section we describe the signal processing chain for ANITA-IV, and in particular
the steps that are relevant to understanding the role of the TUFF boards.
We will note when and where the ANITA-III signal processing differed.
The RF signal processing chain for ANITA-IV is illustrated in Figure~\ref{system}. 
Each ANITA antenna is dual-polarized with feeds for vertically and horizontally polarized (VPol and HPol) signals.
Therefore, for 48 antennas there are 96 total full-band ($200 - 1200\,\mbox{MHz}$) RF signal channels.

Each channel goes through the 
Antenna-Mounted Pre-amplifier (AMPA) unit before entering the Instrument Box. 
There is an AMPA unit connected directly to the VPol and HPol outputs of each antenna.
The AMPA contains a $200 - 1200\,\mbox{MHz}$ bandpass filter, followed by an approximately $35\,\mathrm{dB}$ Low Noise Amplifier (LNA). 
Following the AMPA unit, the RF signal travels through $12\,\mbox{m}$ of
LMR240 coaxial cable to the Instrument Box. 
Inside the Instrument Box, the signal first goes through second-stage amplification (performed by a different module in ANITA-III) 
and notch filtering (unique to ANITA-IV), both performed by the TUFF boards in ANITA-IV.
Then it passes through another set of bandpass filters before
being split into digitization and triggering paths. 
The triggering and digitization processes are detailed below.

\subsection{Triggering:}
\label{trigger}

\begin{figure}[H]
\centering
\subfigure{
	\includegraphics[width=0.93\textwidth]{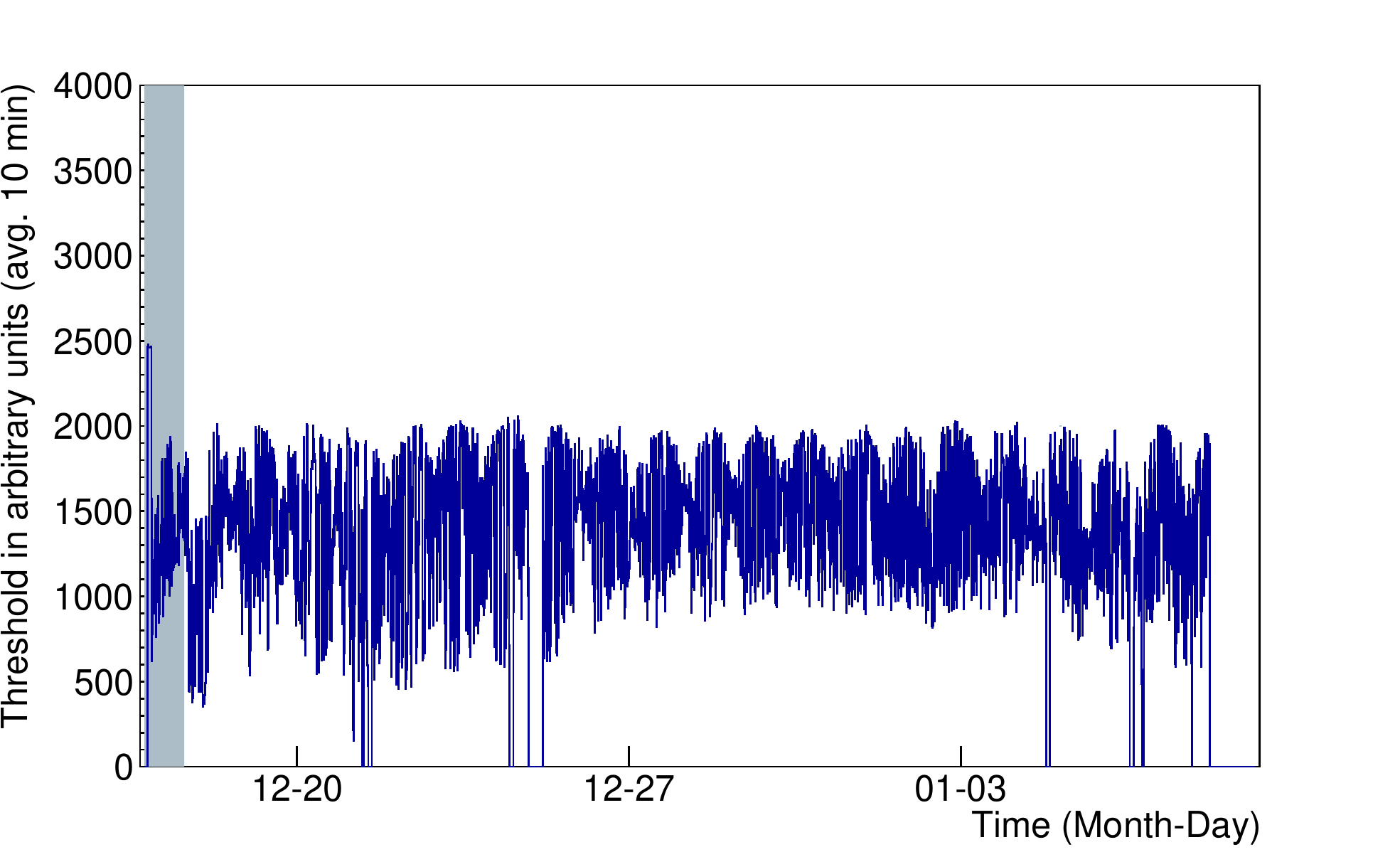}
	\label{anita3thresholds}
}
\subfigure{
	\includegraphics[width=0.93\textwidth]{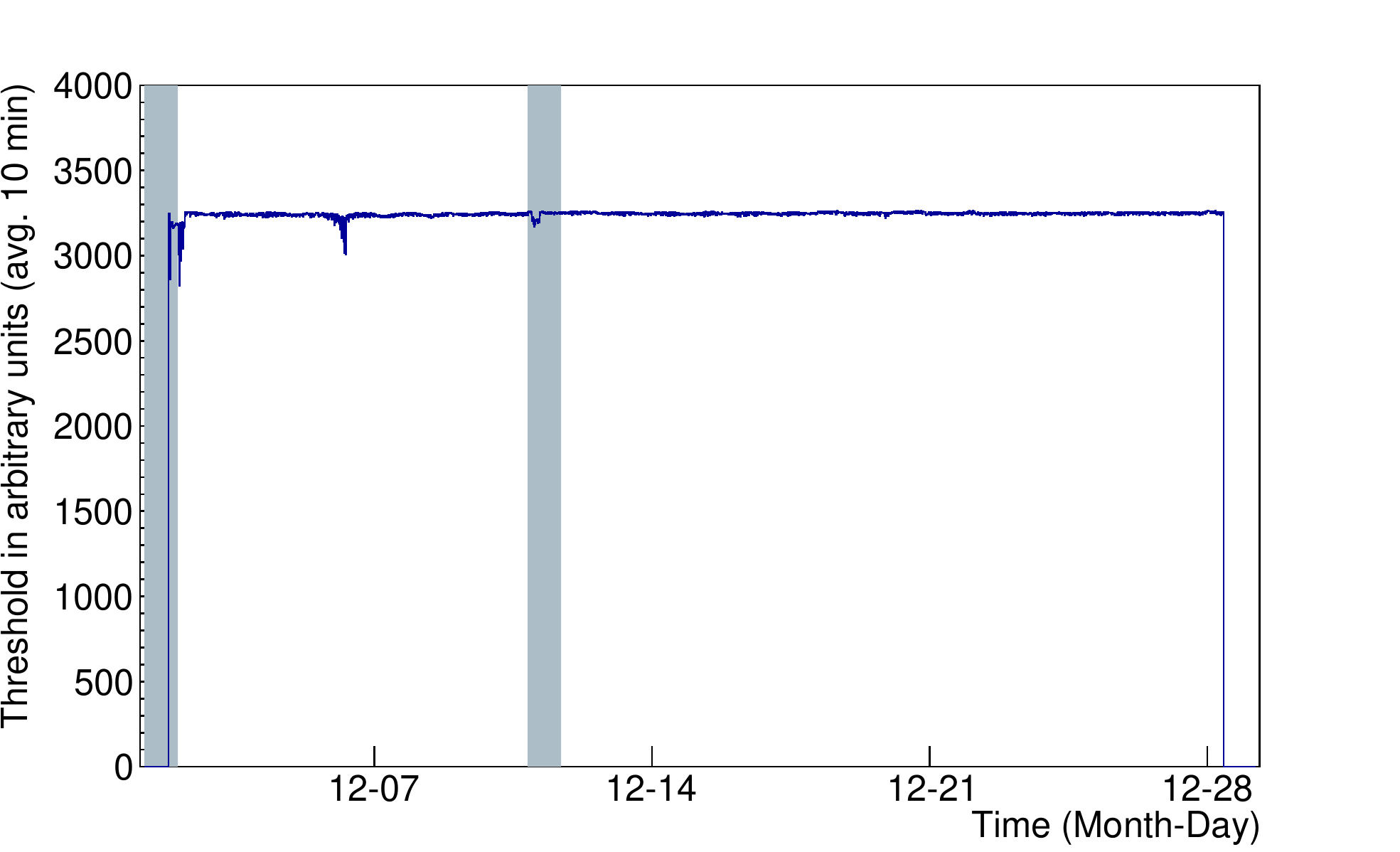}
	\label{anita4thresholds}
}
\caption[]{SURF DAC thresholds in arbitrary units of DAC counts for a single channel for the ANITA-III (top) and ANITA-IV (bottom) flights. 
Changing thresholds is a secondary method of avoiding digitization deadtime due to CW interference. 
The TUFF boards helped to maintain constant thresholds in ANITA-IV, 
whereas in ANITA-III, thresholds had to be changed throughout the flight.
Note that a lower threshold corresponds to a higher and therefore, stricter requirement on the power of the incoming signal, and so during periods of high anthropogenic noise, the SURF DAC thresholds were lowered.
The shaded regions indicate when the 
ANITA payload was in line of sight of the NASA LDB Facility. 
}
\label{thresholds}
\end{figure}

In the triggering path, the RF signals from both the VPol and HPol channels of a single antenna 
are passed through a $90^{\circ}$ hybrid (hybrids were absent in ANITA-III). 
The outputs from the $90^{\circ}$ hybrid are the left and right circular polarized
(LCP and RCP) components of the combined VPol and HPol signals from an antenna. 
The hybrid outputs are input to the SURF (Sampling Unit for RF) high-occupancy RF Trigger (SHORT) unit before being passed to the SURF board. 
Each SHORT takes four channels as its input. 
In a SHORT
channel, the RF signal passes through a tunnel diode and an amplifier. 
The output of the SHORT is
approximately proportional to the square of the voltages
of the input RF signal integrated over approximately $5\,\mbox{ns}$.
It is a measure of the power of the incoming signal and is typically a negative voltage.
The SHORT output is routed to a SURF trigger input where 
it enters a discriminator that compares this negative voltage in Digital-to-Analog Converter (DAC) counts to the output
of a software-controlled DAC threshold on the SURF, henceforth referred to as the SURF DAC threshold. 
The SURF DAC threshold is expressed in arbitrary units of DAC counts corresponding to voltages. Lower thresholds 
correspond to higher voltages and therefore, higher power of the incoming signal. 
The SURF DAC threshold can be changed during flight. 
During the ANITA-III flight, CW interference overwhelmed the digitization system, forcing us to impose
frequent and large changes in the SURF DAC thresholds. 
A comparison of SURF DAC thresholds between ANITA-III and ANITA-IV is presented in Figure~\ref{thresholds}.

\paragraph{Trigger logic:}
Due to power and bandwidth limitations, ANITA is not able to constantly record data. 
Digitization of data only occurs when the trigger conditions are satisfied.
The ANITA-IV trigger consists of three triggering levels: Level~1, Level~2 and Level~3. 
The trigger requirements at each of these three levels is described below.

\paragraph{Level~1 trigger:}

The Sampling Unit for RF (SURF) board issues the Level~1 trigger.
To form a Level~1 trigger, the SHORT outputs of the LCP and RCP channels from the same antenna 
are required to exceed the SURF DAC threshold within $4\,\mbox{ns}$. This LCP/RCP coincidence requirement 
was added to the ANITA-IV trigger to mitigate anthropogenic and thermal backgrounds.
The signals of
interest are known to be linearly polarized, whereas satellite emission is often circularly polarized
and thermal noise is unpolarized. In the presence of a continuous source of CW signal such as satellites,
the LCP/RCP coincidence may still allow a combination of circularly polarized satellite noise and the circularly polarized component of
thermal noise to satisfy the Level~1 trigger requirement. Therefore, the LCP/RCP coincidence aids in
reducing triggers induced by satellites but does not completely mitigate their effect.

\paragraph{Level~2 trigger:} 

The SURF board issues the Level~2 trigger. 
A Level~1 trigger opens up a time window.
If there are two Level~1 triggers in the same phi sector within the allowable time window, then a Level~2 trigger is issued. 
The allowable time window depends on which antenna had the first Level~1 trigger. 
Time windows of
$16\,\mbox{ns}$, $12\,\mbox{ns}$ and $4\,\mbox{ns}$ in duration are
opened up when a Level~1 trigger is issued in the bottom, middle and top ring respectively.
These time windows were chosen to preferentially select signals coming up from the ice. 
The Level~2 trigger decisions are passed from the SURF boards to a dedicated triggering board called the Triggering Unit for RF (TURF).
The Level~2 trigger timing in ANITA-IV differed from that used in ANITA-III as changes were made to further restrict
the allowed timing of the antenna coincidences to 
better match timing expected from an incoming plane wave.

\paragraph{Level~3 trigger:}

The TURF board issues the Level~3 trigger.
A field programmable gate array (FPGA) on the TURF board monitors Level~2 triggers.
A Level~3 trigger is issued by the TURF board when there are Level~2 triggers in two adjacent phi sectors within $10\,\mbox{ns}$. 
When there is a Level~3 trigger, the TURF board instructs 
the SURF board to begin digitization.

\subsection{Digitization:}

The digitization of the signal is performed by the SURF board. 
There are twelve SURF boards with eight signal channels each for the 96 RF channels. 
The SURF board contains four custom-built Application Specific Integrated
Circuits called Large Analog Bandwidth Recorder And
Digitizer with Ordered Readout (LAB). 

\paragraph{LAB chip and digitization deadtime:}
ANITA-IV uses the third generation
of LAB chips that are described by Varner \textit{et al}. \cite{labrador}. 
The RF signal entering the SURF gets split and fed into four parallel LAB chips (forming four ``buffers" for digitization). 
Each LAB chip has nine 260-element switched capacitor arrays (SCAs). 
The SCAs sample waveform data at the rate of $2.6\,\mbox{GSa/s}$. 
At any moment, the charge stored in an SCA is a $100\,\mbox{ns}$ record of the signal voltage. 
This $100\,\mbox{ns}$ snapshot of the incoming plane wave is known as an ``event."
When a Level~3 trigger occurs, a single LAB chip stops sampling and is ``held.'' It then digitizes
the stored data, which is then read out by the flight computer, taking approximately $5-10\,\mbox{ms}$.
If all four LAB chips are held, the trigger is ``dead'' and the accumulated time when the trigger is dead is recorded as
digitization deadtime by the TURF board. 

\paragraph{Masking:}

\begin{figure}[H]
\centering
\subfigure{
	\includegraphics[width=0.93\textwidth]{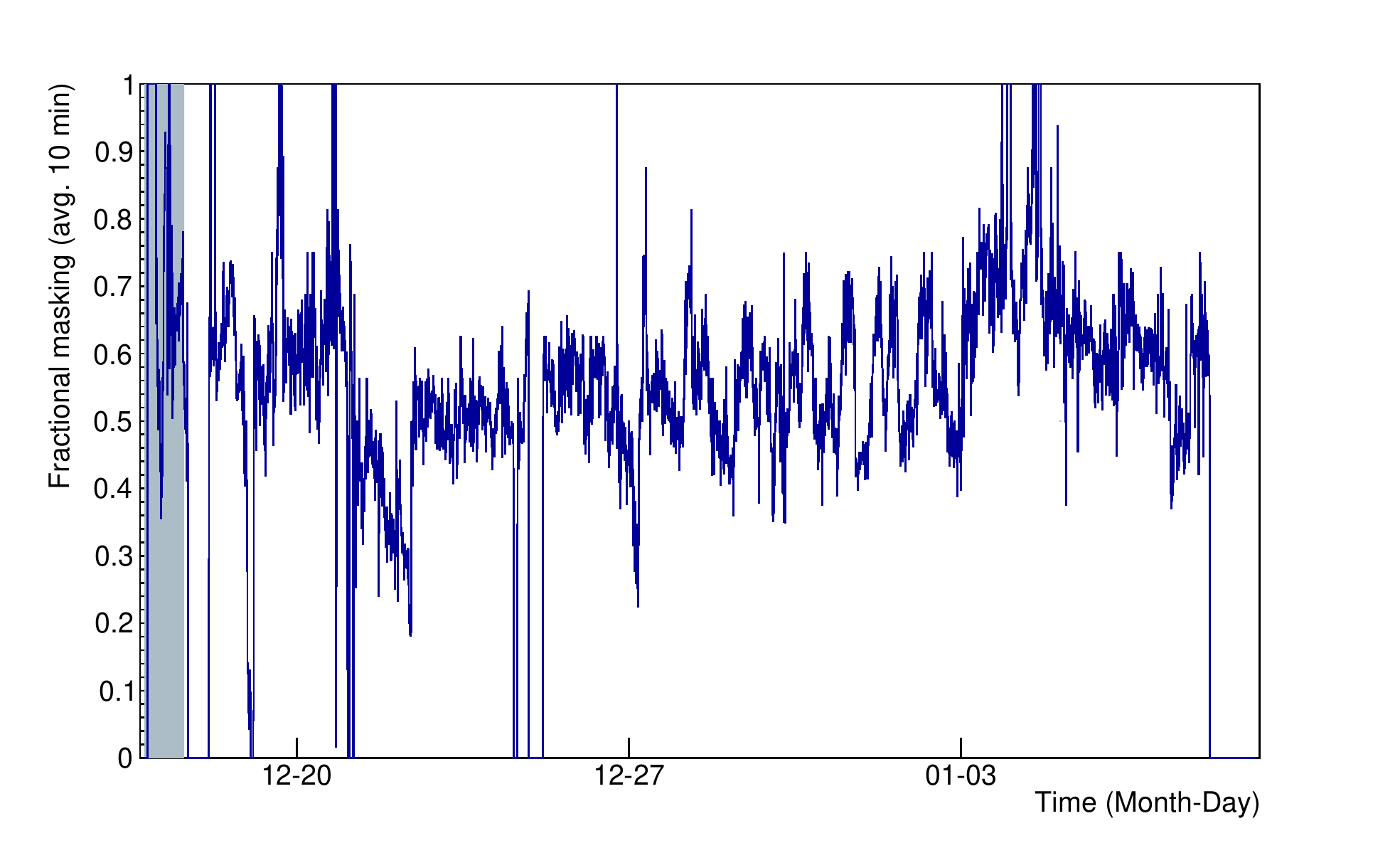}
	\label{}
}
\subfigure{
	\includegraphics[width=0.93\textwidth]{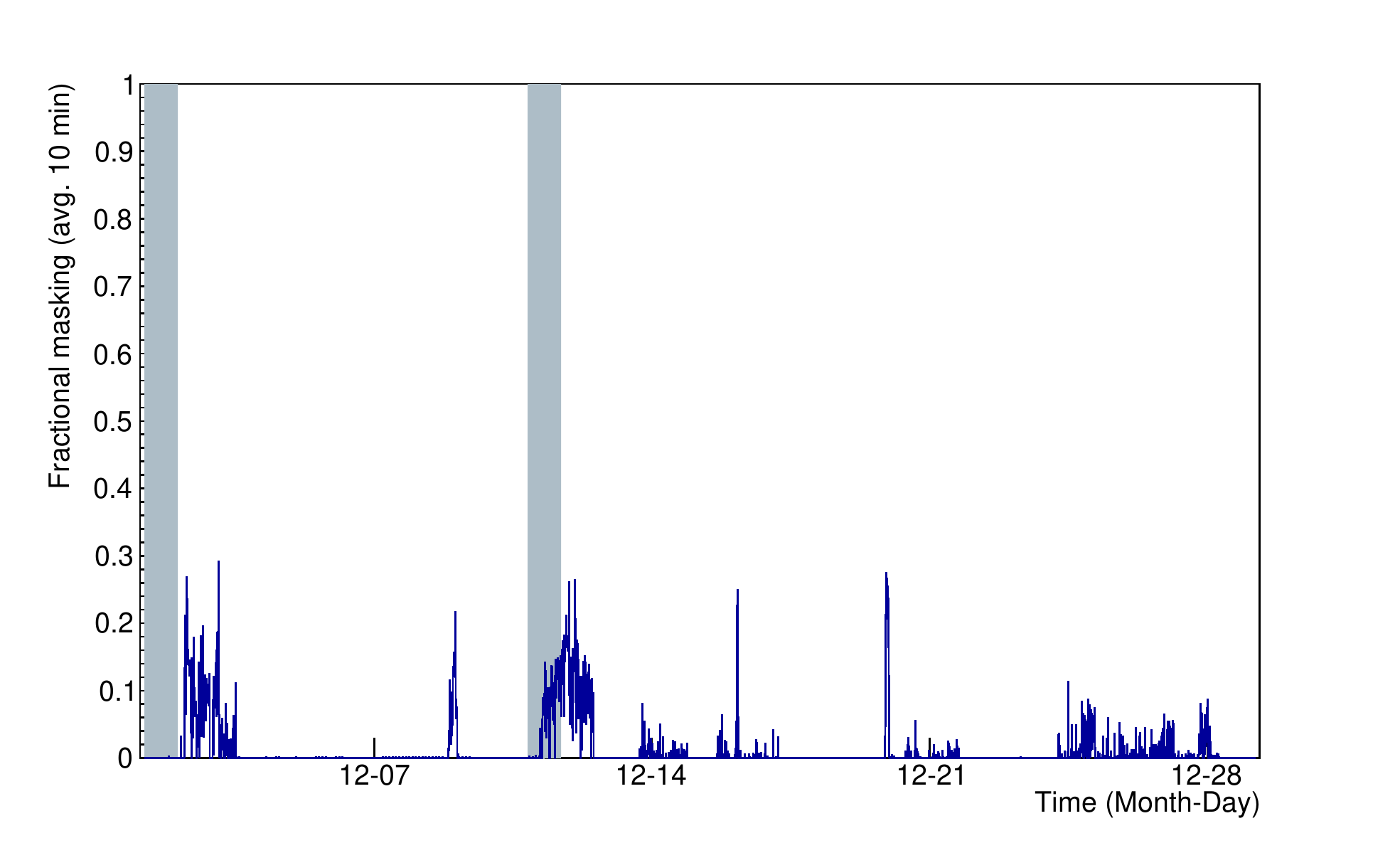}
	\label{}
}
\caption[]{Masking in the ANITA-III (top) and ANITA-IV (bottom) flights. 
Before ANITA-IV, masking was the primary method of avoiding digitization deadtime
due to CW interference. For the majority of the ANITA-III flight, over half of the payload was masked.
Due to the mitigation of CW noise in ANITA-IV to acceptable levels by the TUFF notch filters, the need for masking was strikingly reduced.
The shaded regions indicate when the 
ANITA payload was in line of sight of the NASA LDB Facility. 
}
\label{phimasking}
\end{figure}

During ANITA-III, digitization deadtime due to high levels of anthropogenic noise was reduced
by excluding
certain phi sectors from participating in the Level~3 trigger. 
This is called phi-masking. 
Alternatively, specific channels (each antenna has two channels) were excluded from participating in the Level~1 trigger. 
This is called channel-masking.
Together these are referred to as masking.
Because of CW interference by military communications
satellites, over half of the payload had to be masked
during most of the ANITA-III flight. 
This strongly motivated the creation of the TUFF
boards with tunable, switchable notch filters. 
A comparison of masking between ANITA-III and ANITA-IV 
is presented in Figure~\ref{phimasking}. 

\section{TUFF Board Design}
\label{design} 

For ANITA-IV, we built and deployed 16 TUFF boards (not counting spares) with 
six channels each for the 96 total full-band RF channels of ANITA. 
Figure~\ref{system} shows, for a single RF channel in ANITA-IV, where 
the TUFF boards are in the signal processing chain. 

\begin{figure}[H]
\centering
\includegraphics[width=1.0\textwidth]{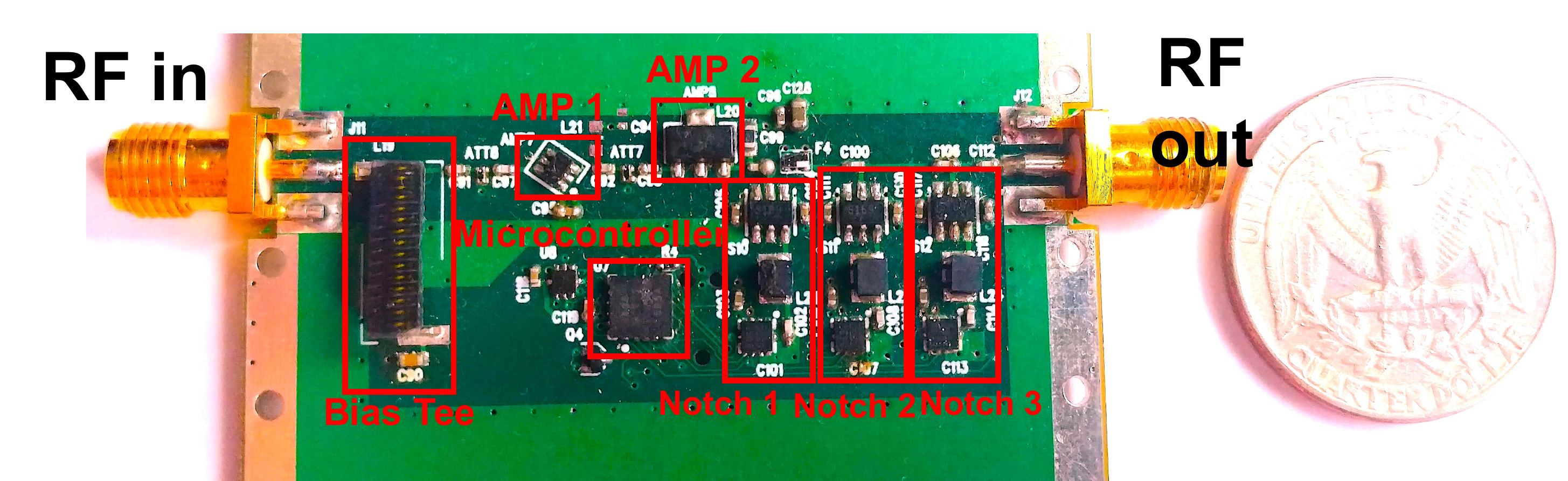}
\caption{A single TUFF board unit (channel) that powers the first-stage antenna-mounted amplification unit 
and performs second-stage amplification and notch filtering of a single RF channel (out of 96 total). 
Each TUFF board has six such channels. The main components of the channel are highlighted here.}
\label{tuff_channel}
\end{figure}

The design of the TUFF board was affected by the low power budget of 
ANITA as well as the weight and size restrictions of a balloon mission, as described in Section~\ref{payload}. 
The TUFF boards needed to be low-power, compact and light. Figure~\ref{tuff_channel} shows a 
single TUFF channel next to a quarter USD coin for size comparison. 
It can be seen that a single channel is about twice the size of the coin. 
Each printed circuit board has four layers of copper with an FR-4 dielectric material. 
The TUFF boards operate on $3.3\,\mathrm{V}$ and $4.7\,\mathrm{V}$ power sources 
provided by
a MIC5504 from Microchip Technologies Inc. and a ADM7171 from Analog Devices Inc. Both
voltage regulators draw from a $5\,\mathrm{V}$ source 
supplied by the DC/DC unit in the ANITA
Instrument Box. 
A single TUFF channel consumes only $330\,\mathrm{mW}$ of power. 
The total power consumed by the ANITA payload is approximately $800\,\mathrm{W}$. 

\begin{figure}[H]
\centering
\includegraphics[width=1.0\textwidth]{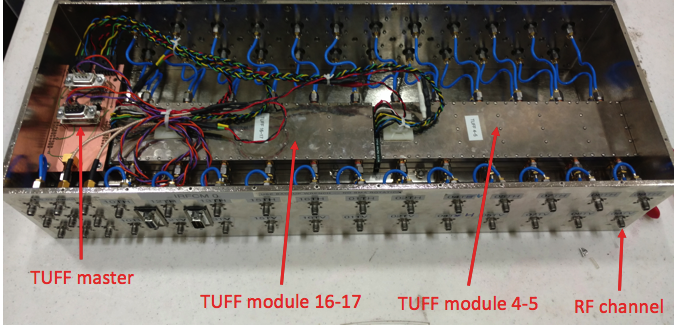}
\caption{Internal Radio Frequency Conditioning Module (IRFCM) containing two 12 channel 
TUFF modules serving 24 RF channels total, together with a TUFF Master for sending commands to the TUFF boards from the flight computer.}
\label{IRFCM}
\end{figure}

Two TUFF boards were assembled into a final 
12-channel aluminum housing. This provides heat-sinking, structural support, and RF isolation. 
Two of these 12-channel modules were placed inside an 
Internal Radio Frequency Conditioning Module (IRFCM) 
inside the 
Instrument Box of ANITA. Figure~\ref{IRFCM} shows the inside of an IRFCM. 

Each TUFF channel has four main components. These are highlighted in 
Figure~\ref{tuff_channel} and described below.

\subsection{Amplifiers and bias tee} 

\begin{figure}[H]
\centering
\includegraphics[width=1.0\textwidth]{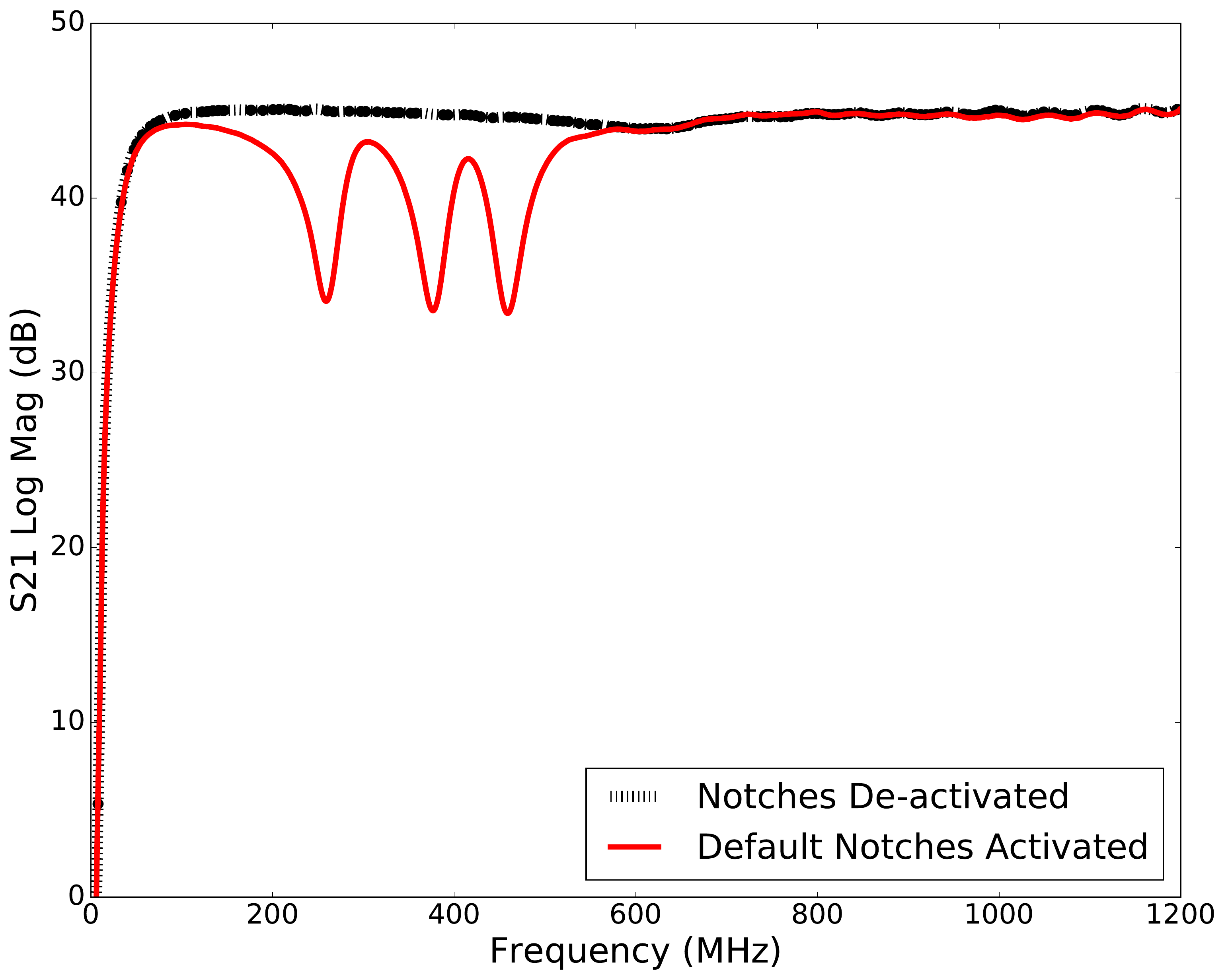}
\caption{The forward transmission coefficient, S21, or the gain of a TUFF channel as measured in the lab with all notches de-activated (black dashed line) and all notches activated at their default frequencies (red solid line). There is approximately $13\, \mathrm{dB}$ of attenuation in the notched regions.}
\label{s21}
\end{figure}

There are two amplifiers connected in series that together 
produce second-stage RF power amplification of approximately $45\,\mathrm{dB}$. 
The gain of a TUFF channel, as measured in the lab, is shown in Figure~\ref{s21}. 
 In Figure~\ref{tuff_channel}, AMP~1 is a BGA2851 by NXP Semiconductors and 
AMP~2 is an ADL5545 by Analog Devices. 
There is an attenuator producing $1\,\mathrm{dB}$ of 
attenuation to the RF signal as it leaves
AMP~1 and before it enters AMP~2.
The BGA2851 provides a gain of $24.8\,\mathrm{dB}$ at $950\,\mbox{MHz}$. 
It has a noise figure~of $3.2\,\mathrm{dB}$ at $950\,\mbox{MHz}$. 
It consumes $7\,\mathrm{mA}$ of current at a supply voltage of $5\,\mathrm{V}$, 
or $35\,\mathrm{mW}$ of power.
The ADL5545 provides a gain of $24.1\,\mathrm{dB}$ with broadband operation from $30-6000\,\mbox{MHz}$.
Out-of-band power at frequencies above $2\,\mbox{GHz}$ is suppressed by a filter on each TUFF channel. 
Additionally, there are band-pass 
filters immediately after the TUFF boards in the signal processing chain allowing power only in the frequency range $200 - 1200\,\mbox{MHz}$. 
The ADL5545 has a noise figure~of $2.9\,\mathrm{dB}$ at $900\,\mbox{MHz}$ 
and a $1\,\mathrm{dB}$ compression point (P1dB) of $18.1\,\mathrm{dBm}$ at $900\,\mbox{MHz}$. 
It consumes $56\,\mathrm{mA}$ of current at a supply voltage of $5\,\mathrm{V}$, or $300\,\mathrm{mW}$ of power. 
Thus, this amplifier consumes the majority of the power required by a single TUFF channel. 

There is a bias tee on each TUFF channel that 
remotely powers the AMPA (antenna-mounted pre-amplifier) unit at the other end of the coaxial cable connecting an AMPA and that channel. 
It consists of a 4310LC inductor by Coilcraft in series with a $0.1\,\mathrm{\mu F}$ capacitor. 
The inductor delivers DC to the AMPA unit while the capacitor prevents DC from passing through to the signal path of the TUFF channel. 
The bias tee allows RF signal traveling from the AMPA unit through the coaxial cable to pass 
through to the rest of the signal path of the TUFF channel. 

\subsection{Notch filters} 

There are three tunable, switchable notch filters 
for mitigation of CW noise at the default frequencies of $260\,\mbox{MHz}$ (Notch~1), $375\,\mbox{MHz}$ (Notch~2) 
and $460\,\mbox{MHz}$ (Notch~3).
The gain of a TUFF channel, with all notch filters activated, is shown in Figure~\ref{s21}. 
The TUFF notches were able to achieve a maximum attenuation of approximately $13\, \mathrm{dB}$, and were implemented as a simple RLC trap, with the resistance $R$
originating from the parasitic on-resistance of a dual-pole, single-throw
RF switch and the DC resistance of the remaining components. This is approximately $6 - 7\,\mathrm{\Omega}$. 
The inductance $L$ is fixed at $56\,\mathrm{nH}$. The capacitance $C$ is a
combination of a fixed capacitor and a PE64906 variable capacitor from
Peregrine Semiconductor. Simulations using the device model of the
variable capacitor suggest that the mounting pads of the components
contribute $\sim~0.6\,\mbox{pF}$ of parasitic capacitance.

\begin{figure}[H]
\centering
\includegraphics[width=1.0\textwidth]{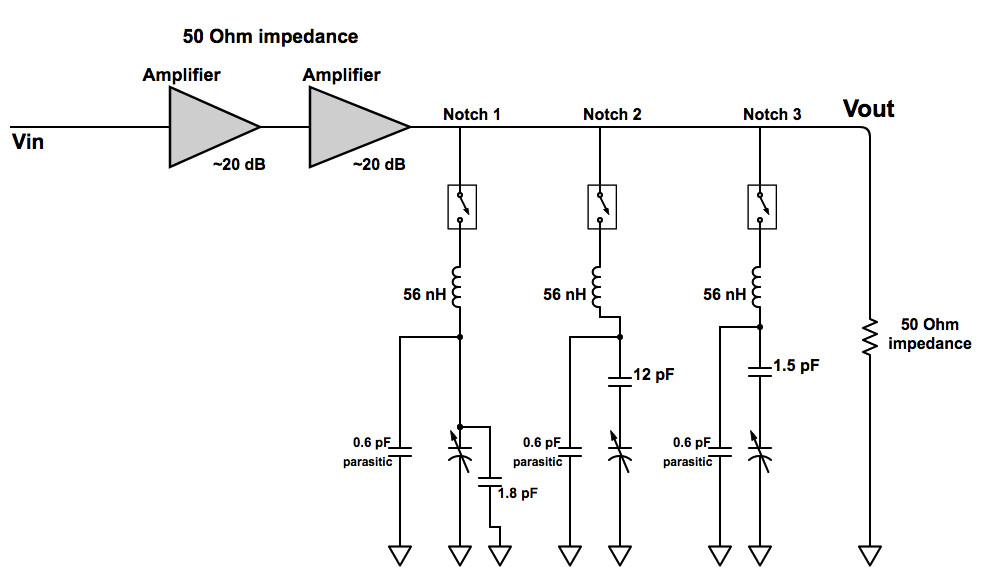}
\caption{Circuit diagram showing the different components of the TUFF notch filters.}
\label{circuit}
\end{figure}

With the 
tuning capability of the variable capacitor, the resonant frequency of the RLC circuit was 
modified during flight to dynamically mitigate CW interference. 
The variable capacitor in a notch can be 
tuned in 32 discrete steps of $119\,\mathrm{fF}$ in the range $0.9-4.6\,\mathrm{pF}$ and for 
each notch, is connected in series or parallel with a constant capacitance. 
For Notch~1, the variable capacitor is 
in parallel with a $1.8\,\mathrm{pF}$ capacitor. For Notches~2 and 3, the variable capacitor is in 
series with a $12.0\,\mathrm{pF}$ (Notch~2) and a $1.5\,\mathrm{pF}$ (Notch~3) capacitor for 
increased tuning capability. Figure~\ref{circuit} shows a simplified circuit diagram.

\subsection{Microcontroller}

We use an ultra-low-power microcontroller, specifically a MSP430G2102 by Texas Instruments.
This features a powerful 16-bit Reduced Instruction Set Computing (RISC) central processing unit (CPU). 
There are five low-power modes optimized for extended battery life. 
The active mode consumes $220\,\mu\mbox{A}$ at $1\,\mbox{MHz}$ and $2.2\,\mathrm{V}$. 
The standby mode consumes only $0.5\,\mu\mbox{A}$ and the RAM retention-off mode consumes $0.1\,\mu\mbox{A}$.
The digitally-controlled oscillator allows wake-up from low-power modes to active mode in less than 
$1\,\mu\mbox{s}$. 

During the ANITA-IV flight, commands could be sent using the SIP connection to set the 
state of the variable capacitor of each TUFF notch filter via the microcontroller 
of that channel. 
This was done in real time if a re-tune of a notch filter was necessary to mitigate CW interference.
Commands could be sent to de-activate or activate a notch filter using the switch associated with each notch. 
Each microcontroller has the capability to communicate over universal serial communication interface.

\section{TUFF notch filter operations during the ANITA-IV flight}

\begin{figure}[H]
\centering
\includegraphics[width=1.0\textwidth, height=0.15\textheight]{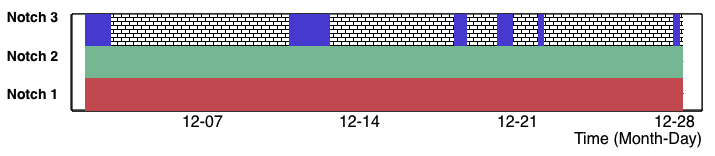}
\caption{The activated (solid red for Notch~1, solid green for Notch~2, solid blue for Notch~3) 
or de-activated (hatched) status for each TUFF notch filter during the flight. 
}
\label{notches_time}
\end{figure}

The TUFF boards were deployed for the first time in
ANITA-IV and 
proved to be critical to the success of the mission. The TUFF notch filters were 
heavily used throughout the flight. 
Figure~\ref{notches_time} summarizes the status of each notch as a function of time during the flight. 

\paragraph{Notch~1: $\bold{260}\,\mbox{MHz}$}

\begin{figure}[H]
\centering
\includegraphics[width=1.0\textwidth]{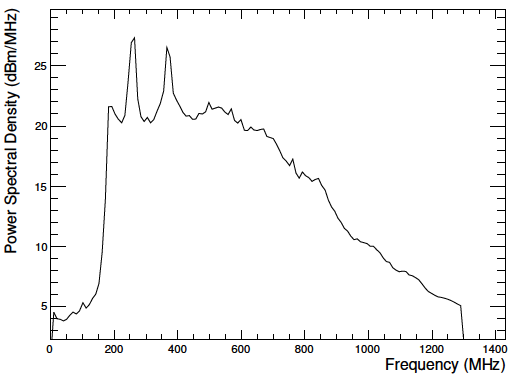}
\caption{A plot of the average power spectral density over $1\,\mbox{min}$ for one channel from 
when the ANITA-III payload was near WAIS Divide in Antarctica. The two peaks at $260\,\mbox{MHz}$ and
 $375\,\mbox{MHz}$, presumably from military satellites, are visible here. The $260\,\mbox{MHz}$ peak 
 was present throughout the flight and the $375\,\mbox{MHz}$ peak was present during less than half of the flight. 
 These CW peaks 
 motivated the installation of the TUFF notch filters in ANITA-IV. As it turns out, Notch~1 (to curb the left peak) and 
 Notch~2 (to curb the right peak) both needed to be active for essentially the entire flight in ANITA-IV.}
\label{anita3spectra_wais}
\end{figure}

During the ANITA-III flight, 
a CW signal at $260\,\mbox{MHz}$ from military satellite systems 
(CW peak seen in Figure~\ref{anita3spectra_wais}) was present throughout the flight. 
This CW signal was omnipresent during the ANITA-IV flight as well, and so Notch~1 needed to be 
active throughout the flight. Notch~1 (usually centered at $260\,\mbox{MHz}$) was re-tuned on Dec 14 as we saw CW interference at 
$250\,\mbox{MHz}$ and was tuned back to $260\,\mbox{MHz}$ later that day. 

 \paragraph{Notch~2: $\bold{360 - 390}\,\mbox{MHz}$}

\begin{figure}
\centering
\includegraphics[width=1.0\textwidth]{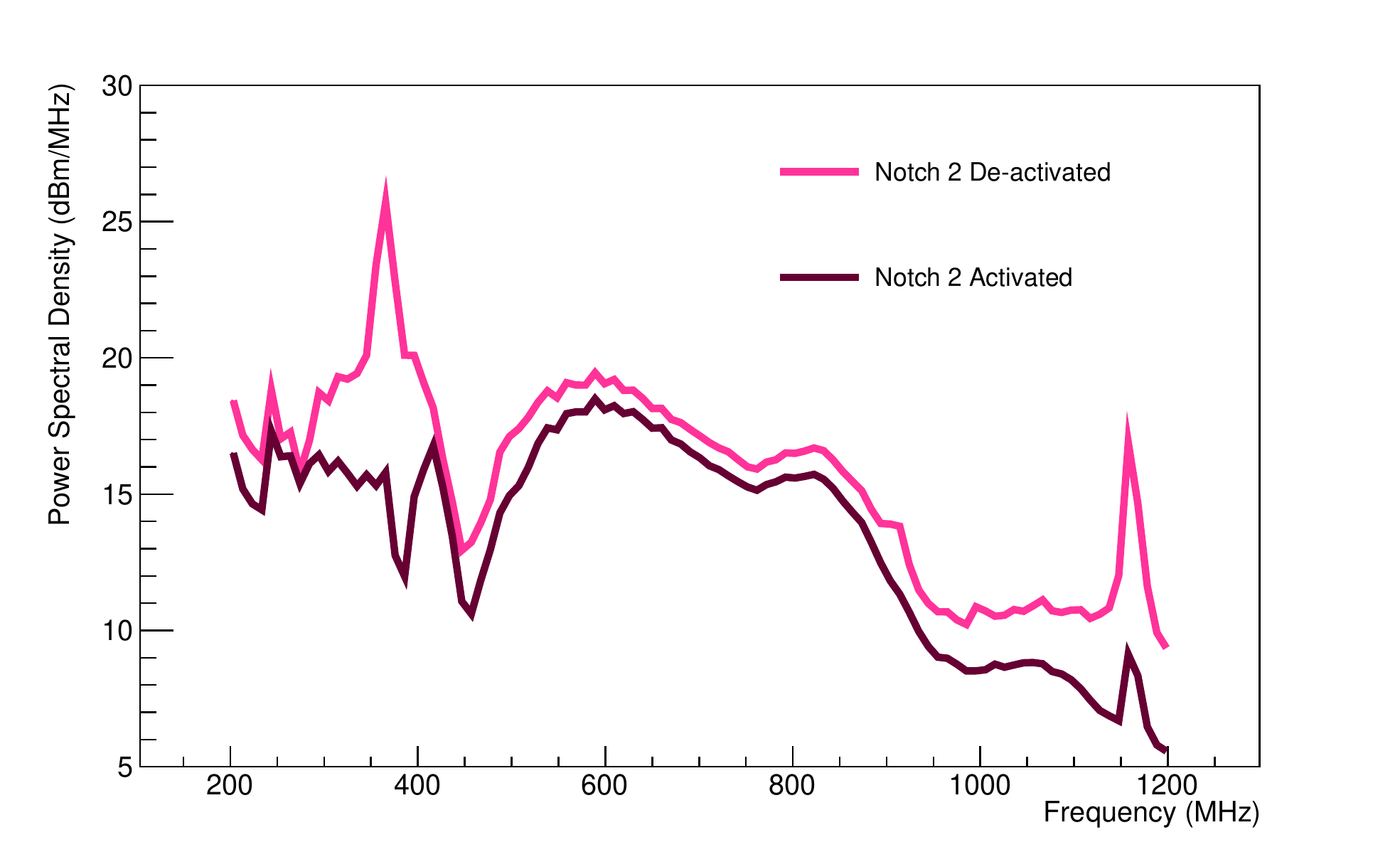}
\caption[]{Power spectra with Notch~2 de-activated and Notch~2 activated (spectra averaged over 16.5 minutes) 
during the ANITA-IV flight. Notch~2 was de-activated on Dec~2 for 16~minutes resulting in a CW peak seen in the spectra. 
Notch~2 was then activated again, and the CW peak was curbed. Although we show only phi sector 16 here, excess 
CW noise upon de-activating Notch~2 and the effect of activating Notch~2 again was seen in almost all phi sectors.}
\label{spectra_notch2_offon_dec2}
\end{figure}

\begin{figure}
\centering
\includegraphics[width=1.0\textwidth]{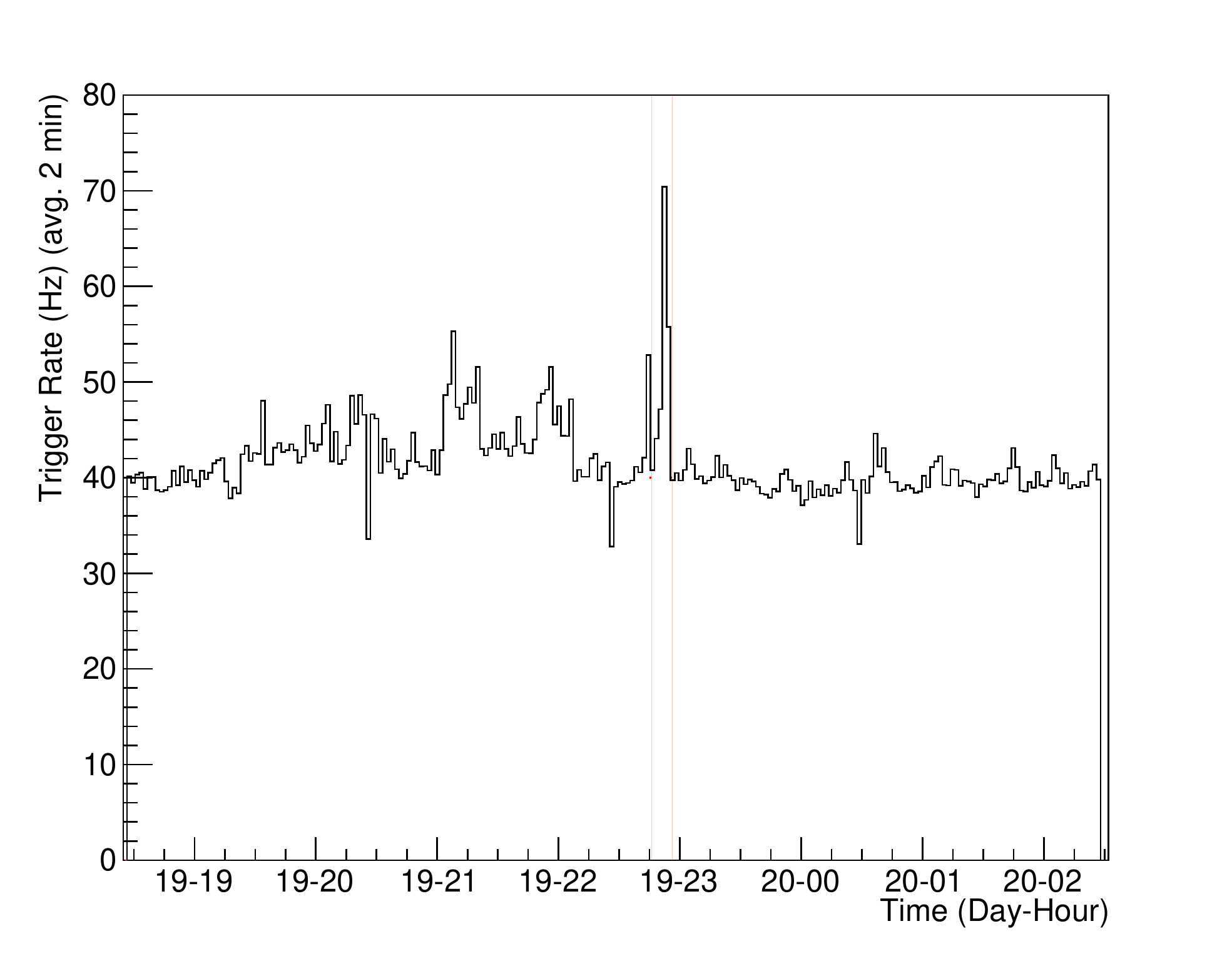}
\caption[]{On Dec~19 at approximately 10:46 PM, Notch~2 was de-activated for approximately 10~minutes. The vertical red lines enclose the duration of time during which Notch~2 was de-activated. 
A trigger rate above $\sim50\,\mbox{Hz}$ incurs digitization deadtime.
The spike in event rate shows that Notch~2 was crucial to keeping CW interference in check. 
Even with the LCP/RCP coincidence required by the ANITA-IV trigger, further mitigation of CW interference by the TUFF boards was necessary to avoid masking.
}
\label{rate}
\end{figure}

\begin{figure}[H]
\centering
\includegraphics[width=1.0\textwidth]{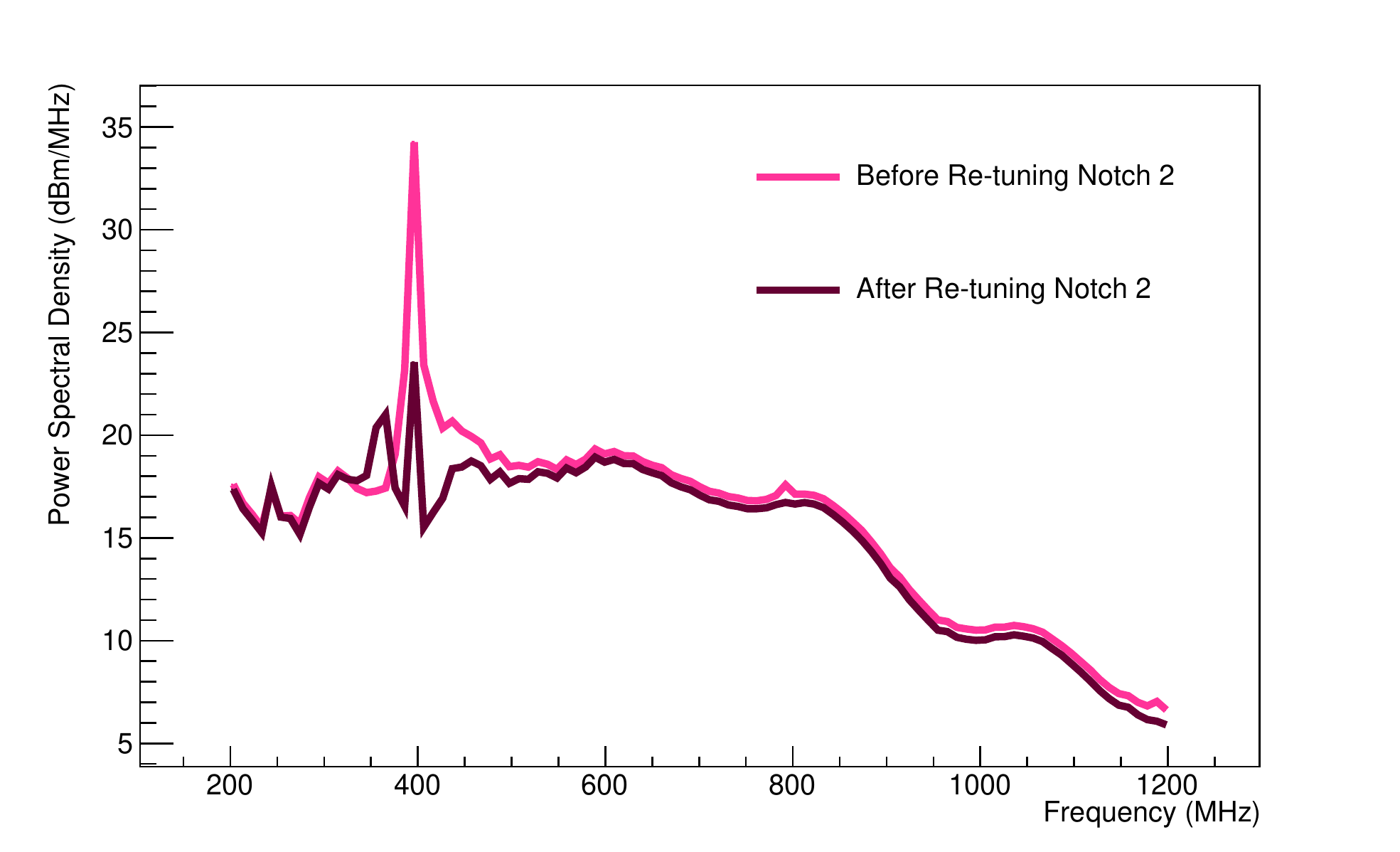}
\caption{Power spectra (averaged over 2 hours) from before and after dynamically tuning 
Notch~2 during the ANITA-IV flight. On observing a large CW peak at $390\,\mbox{MHz}$ on Dec 7, Notch~2 was re-tuned. 
Although we show only phi sector~8 here, similar CW peaks and effects of notch tuning were seen in all phi sectors.}
\label{spectra_notch2_tuning}
\end{figure}

During the ANITA-III flight, a second CW peak at $375\,\mbox{MHz}$ from military satellite systems
 (CW peak seen in Figure~\ref{anita3spectra_wais}) was sometimes present. 
 The MUOS-1 and MUOS-2 satellites are suspected to have caused the second CW peak in ANITA-III.
 This peak is always present during the ANITA-IV flight. 
 The enhanced second peak in ANITA-IV is likely due to the presence of 
 three additional MUOS satellites, that is, MUOS-3, MUOS-4 and MUOS-5, in orbit during the ANITA-IV flight. 
  During the ANITA-IV flight, Notch~2, although de-activated twice (Dec~2, Dec~19), needed 
  to be activated again within minutes due to this CW noise. 
  After being de-activated on Dec~2 for $\sim16\,\mbox{minutes}$, Notch~2 
  needed to be activated again. Excess CW noise upon de-activating Notch~2 was seen in almost all phi sectors. 
  In Figure~\ref{spectra_notch2_offon_dec2} we show spectra averaged over all waveforms from one phi sector during this period. 
  Notch~2 was de-activated again on Dec~19 for approximately 10 minutes. 
  The trigger rate was nearly doubled almost as soon as the notch was de-activated (see Figure~\ref{rate}) and excess CW noise was seen in several phi sectors. 
  Notch~2 was re-tuned during flight a few times (Dec 6-8) to dynamically combat CW interference in 
the range of $360 - 390\,\mbox{MHz}$. Figure~\ref{spectra_notch2_tuning} shows the effect of real time
 tuning of Notch~2 on Dec 7 for mitigation of CW interference at $390\,\mbox{MHz}$. Tuning the notch brought the CW noise power down. 

\paragraph{Notch~3: $\bold{460}\,\mbox{MHz}$}

Notch~3 was generally activated when the payload was in view of Antarctic bases and filtered 
the $450 - 460\,\mbox{MHz}$ frequency region. Notch~3 was de-activated on Dec~2 for a few 
minutes but had to be activated again as the payload was close to McMurdo Station at the time. 

\section{Performance of ANITA-IV compared to ANITA-III}

The TUFF notch filters contributed to a superior handling of CW interference in ANITA-IV compared to ANITA-III. 
During the ANITA-IV flight, we achieved
decreased masking, increased stability of trigger rate and SURF DAC thresholds, and increased instrument livetime. 
These results are summarized in 
Figures~\ref{thresholds}, \ref{phimasking}, \ref{dig_livetime} and \ref{livetime}.

\subsection{Livetime in ANITA} 
\label{livetime_section}

Increasing livetime was the primary motivation behind building and deploying the TUFF boards in ANITA-IV. 
There are two types of livetime in ANITA, which are described below.

\paragraph{Digitization livetime} 

In ANITA, deadtime due to digitization by all four LAB chips of the SURF board is recorded by 
the TURF board, as illustrated in Figure~\ref{system}. 
This deadtime is recorded as a fraction of a second. Digitization livetime per second can be 
obtained by subtracting this from one. 
Increasing the digitization livetime increases the probability of receiving RF signal due to an UHE neutrino. 

\paragraph{Instrument livetime} 

At any given time, the digitization livetime multiplied by the fraction of unmasked phi sectors (after accounting for channel-masking) gives us the instrument livetime per second. 
In other words, instrument livetime accounts for the fraction of observable ice in azimuth after accounting for masking. 

\subsection{Methods adopted to reduce digitization deadtime}
\label{methods}

\paragraph{Masking} 
Before ANITA-IV, the primary method of reducing digitization 
deadtime due to CW signal was masking, which includes both phi-masking and channel-masking.
However, masking leads to instrument deadtime as parts of the payload 
become unavailable for neutrino detection. 
Due to the TUFF boards, fractional masking below $0.3$ was maintained during the ANITA-IV flight,
as seen in Figure~\ref{phimasking}. 

\paragraph{Changing SURF DAC thresholds} In addition to masking, adjusting the SURF DAC thresholds is also a method of reducing digitization deadtime.
The distribution of SURF DAC thresholds for the ANITA-III and ANITA-IV flights is shown in Figure~\ref{thresholds}. 
It is evident that the method of changing thresholds to minimize digitization deadtime 
was heavily 
adopted during the ANITA-III flight.
As the ANITA-III payload was continuously exposed to CW interference, it was unable to maintain stable SURF DAC thresholds. 
As the TUFF boards mitigated CW interference to acceptable levels in ANITA-IV, the thresholds are kept nearly constant during this flight. 

\subsection{Livetime in ANITA-IV compared to ANITA-III} 

\begin{figure}[H]
\centering
\subfigure{
	\includegraphics[width=1.0\textwidth]{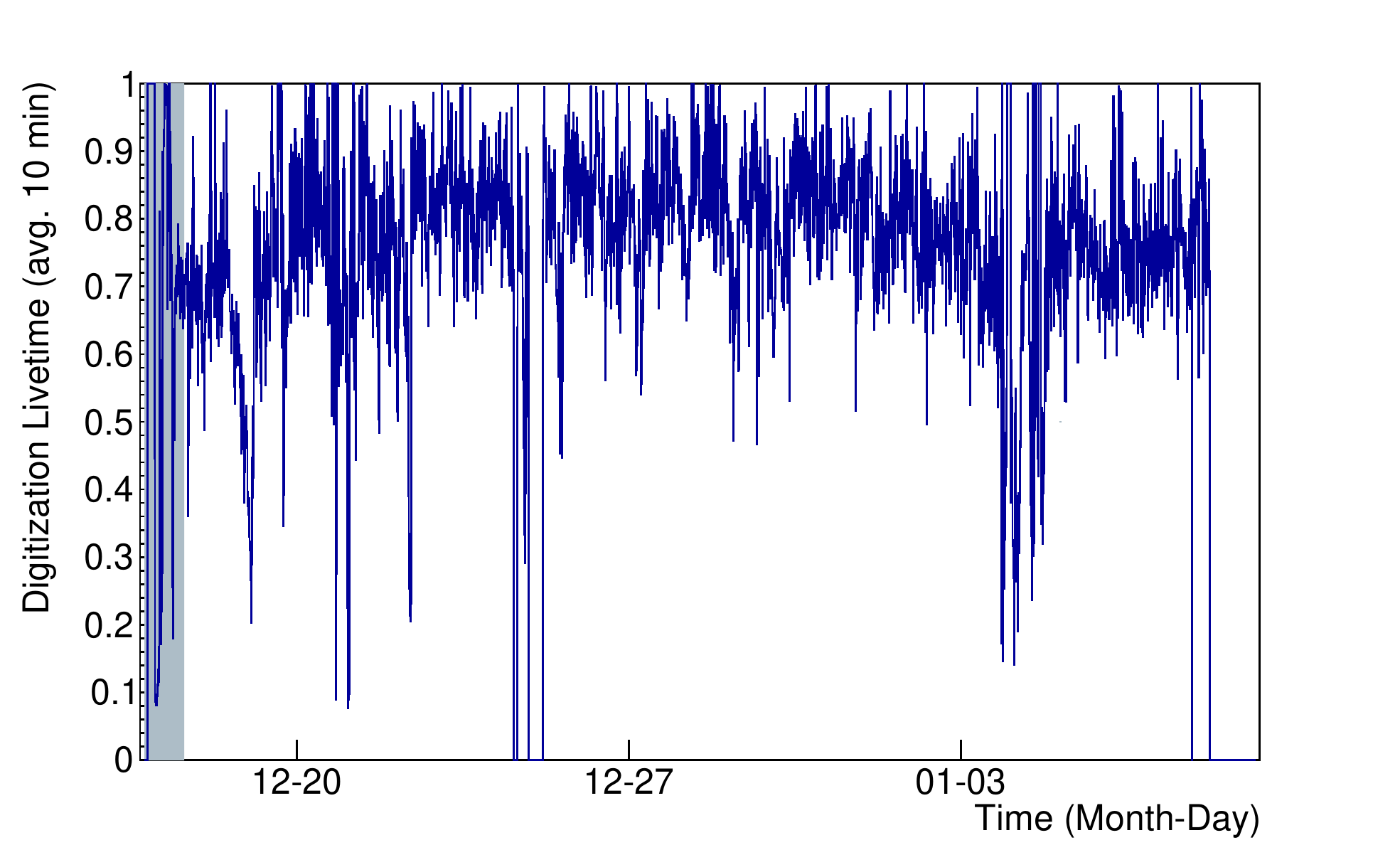}
	\label{anita3livetime}
}
\subfigure{
	\includegraphics[width=1.0\textwidth]{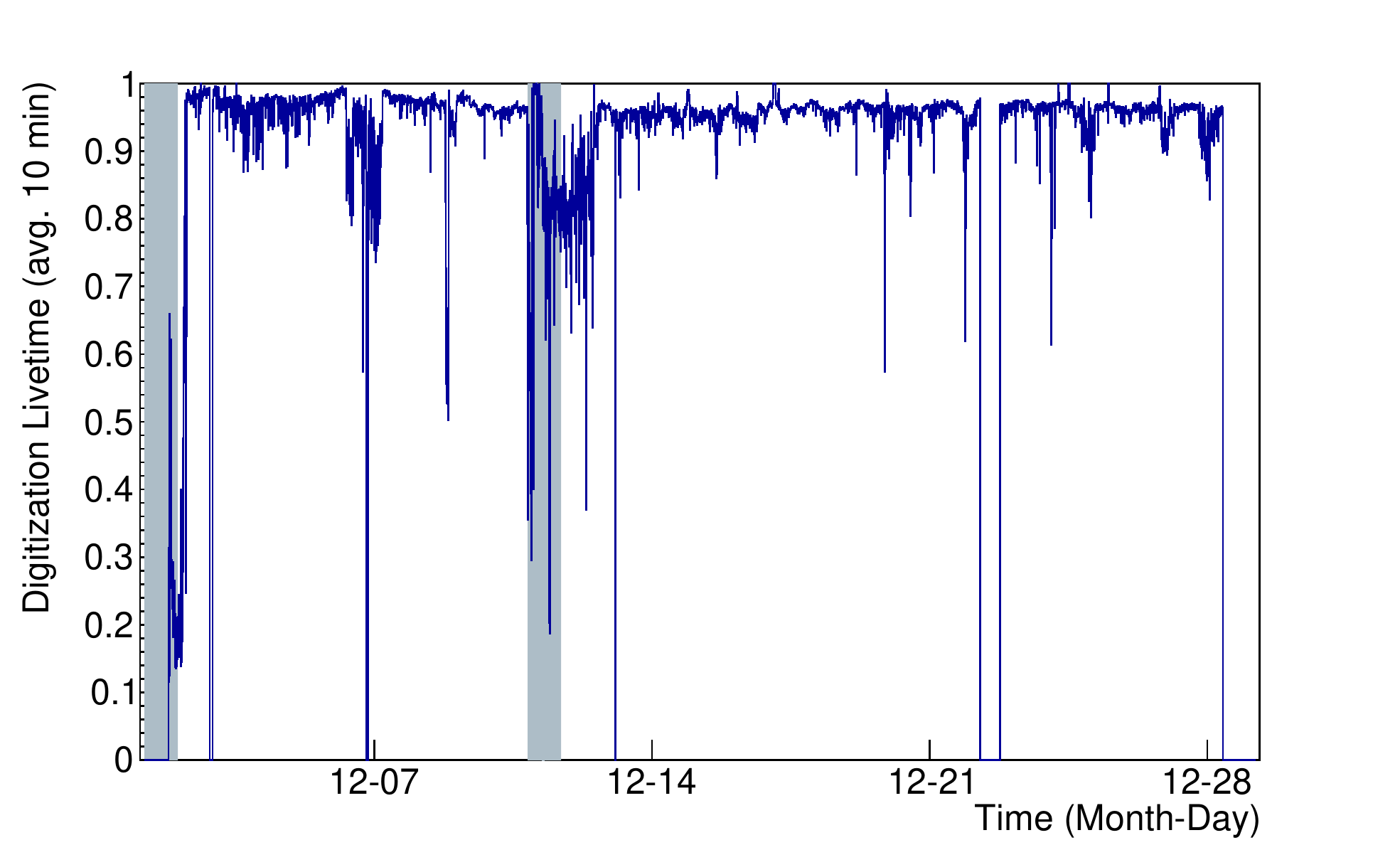}
	\label{anita4livetime}
}
\caption[]{Digitization livetime per second for ANITA-III (top) and ANITA-IV (bottom).
As the ANITA-III payload was inundated by CW interference, digitization livetime was reduced. In ANITA-IV, the TUFF boards helped 
to reduce triggers due to CW signal and therefore, increased the digitization livetime. 
The total digitization livetime for the ANITA-III and ANITA-IV flights 
was calculated to be 73.7\% and 92.3\% respectively.
The shaded regions indicate when the 
ANITA payload was in line of sight of the NASA LDB Facility. 
}
\label{dig_livetime}
\end{figure}

The total digitization livetime for the ANITA-III and ANITA-IV flights was calculated to be 73.7\% and 92.3\% respectively. 
The distribution of digitization livetime per second as a function of time is shown for ANITA-III and ANITA-IV in Figure~\ref{dig_livetime}. 
The TUFF boards dynamically notch-filtered CW peaks in the power spectrum 
of a received signal at an early stage of signal processing.
This 
brought the rate of triggers due to CW signal to acceptable levels and thereby 
increased digitization livetime. 

\begin{figure}[H]
\centering
\subfigure{
	\includegraphics[width=0.97\textwidth]{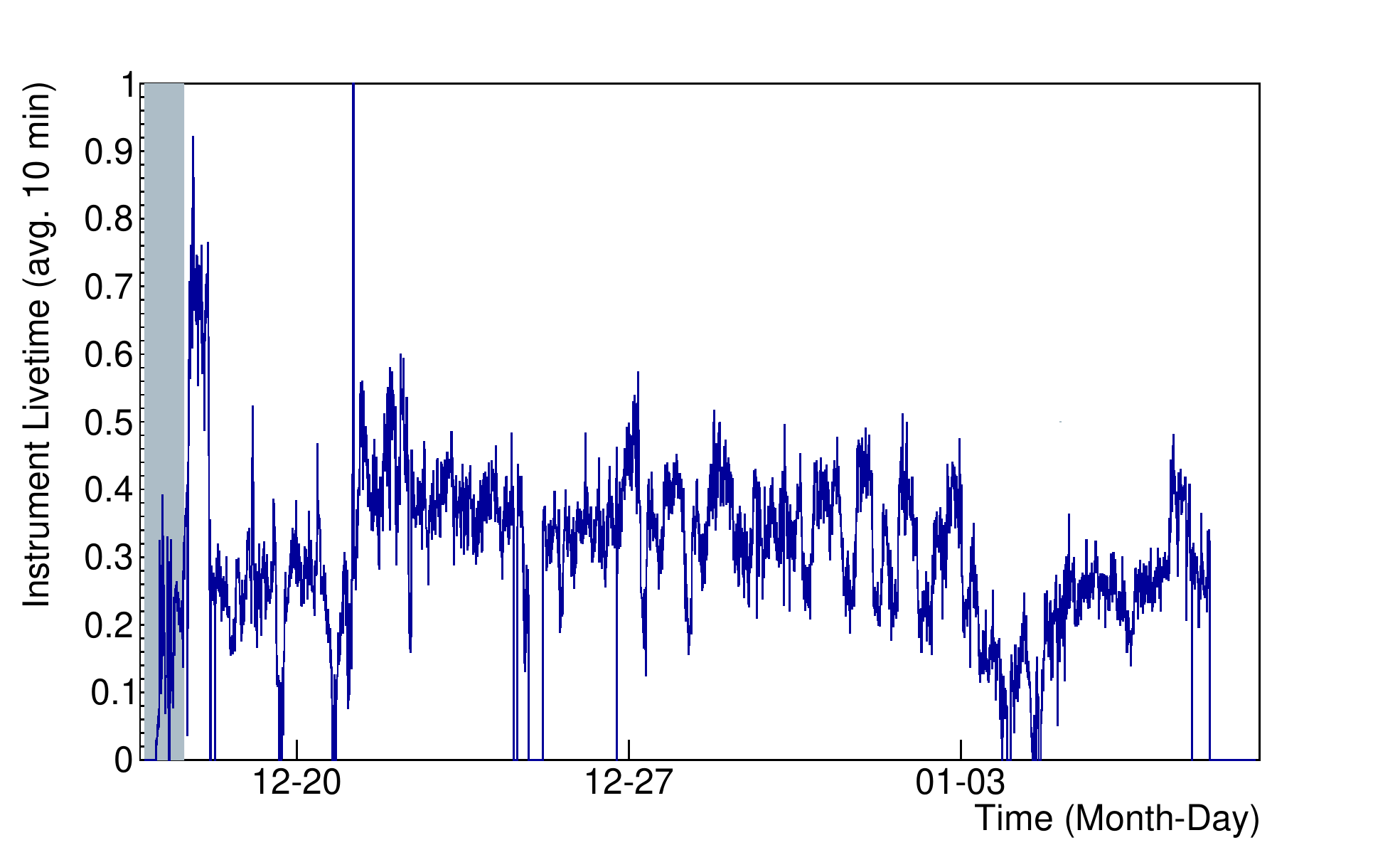}
	\label{anita3livetime}
}
\subfigure{
	\includegraphics[width=0.97\textwidth]{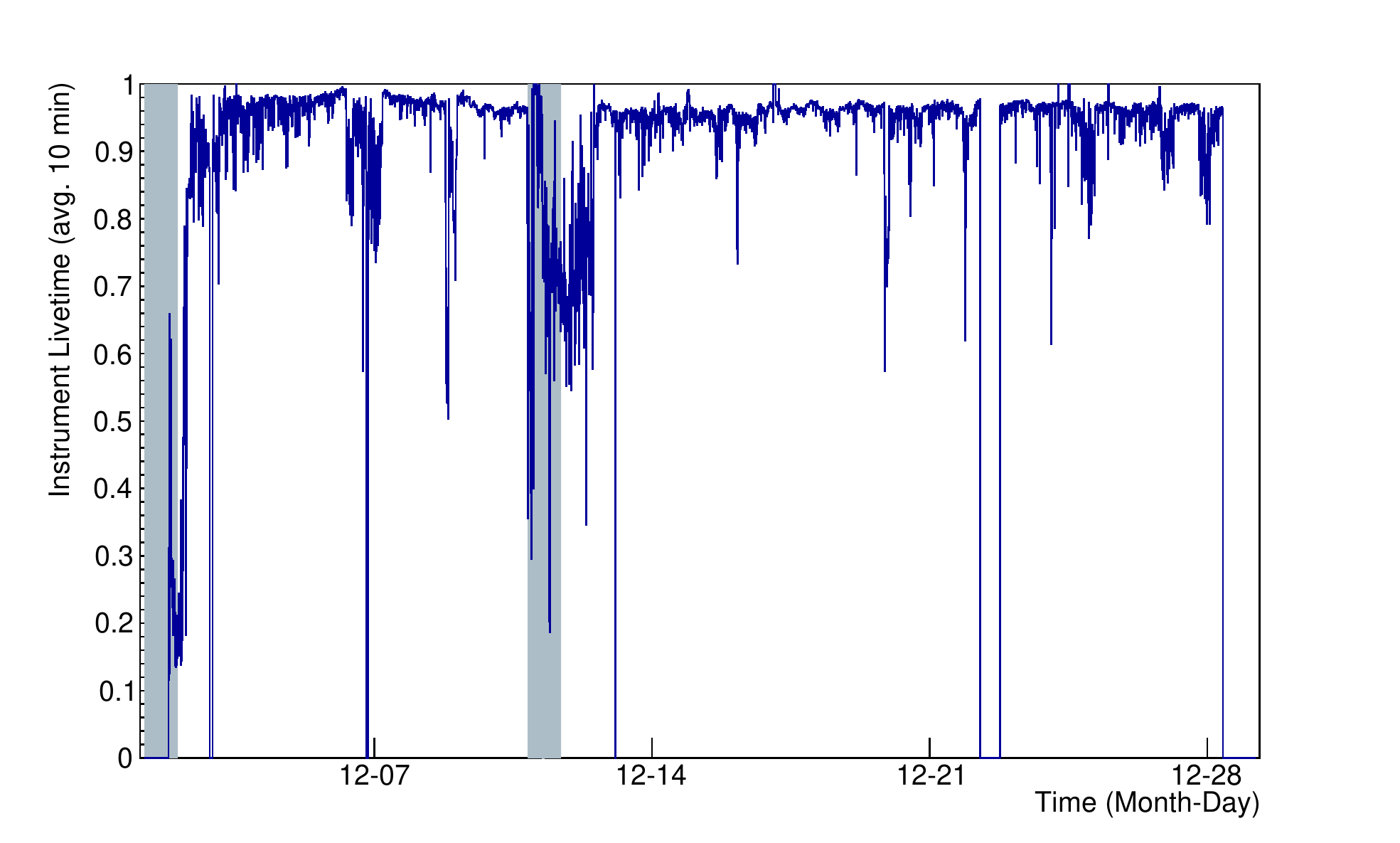}
	\label{anita4livetime}
}
\caption[]{Instrument livetime per second, obtained by weighting digitization livetime by the fraction of
unmasked phi sectors, for ANITA-III (top) and ANITA-IV (bottom). 
In ANITA-III, masking had to be implemented heavily and throughout the flight, which led to a dramatic reduction of
instrument livetime. The TUFF boards largely removed the need for masking in ANITA-IV. 
This helped to increase the 
instrument livetime of ANITA, with 91.3\% total instrument livetime in ANITA-IV,  compared to 31.6\% in ANITA-III.
The shaded regions indicate when the 
ANITA payload was in line of sight of the NASA LDB Facility. 
}
\label{livetime}
\end{figure}

Most importantly, the TUFF boards helped to increase the instrument livetime (digitization livetime weighted by the fraction of
unmasked phi sectors) in the ANITA-IV flight, mainly by decreasing the need for masking. 
The distribution for instrument livetime per second as a function of time is shown for ANITA-III and ANITA-IV in Figure~\ref{livetime}. 
The total instrument livetime for ANITA-III and ANITA-IV was calculated to be 31.6\% and 91.3\% respectively. 
On average, instrument livetime in ANITA-IV was 2.8 times higher than that in ANITA-III. 

\subsection{Impact on signal power and acceptance}

A full account of the impact of the TUFF notch filters on neutrino sensitivity is under investigation
 and beyond the scope of this paper.  We note that each notch removes approximately 5\% 
 of the system bandwidth ($200 - 1200\,\mbox{MHz}$).  Although the impact of increased digitization 
 livetime is straightforward to estimate, the increase in sensitivity
 due to the reduction in masking will require a full account of the time- and azimuthal-dependent exposure of ANITA to neutrinos.

\section{Plans for ANITA-V}

For ANITA-V, we will explore the option of developing software to dynamically activate 
the notch filters only when satellites come into view of an antenna in order to
eliminate any sensitivity loss.
Having established a mechanism for the mitigation of CW interference to acceptable levels, we can now focus 
on improving other parts of the signal processing chain. 
Development of new triggering and digitization systems for ANITA-V is currently underway.
The main proposed upgrade to the triggering system include the Realtime Independent Three-bit Converter (RITC) 
as described by Nishimura \textit{et al}. \cite{ritc}. 
The RITC will perform continuous, low-resolution digitization in order to carry out interferometry of all incoming data in realtime. 
This will be used to generate a system trigger. 
Once triggered, high-resolution digitization of the data will be performed by new SURF boards. 

Proposed upgrades to the SURF board include
new LAB chips (LAB4D), as described by Roberts \textit{et al}. \cite{lab4d}.
There will be 12 LAB chips per SURF board. 
Each LAB chip will sample data from one RF channel using 32 blocks of 128-element SCAs. 
The SCAs will sample waveform data at $3.2\,\mbox{GSa/s}$, eight blocks at a time (forming four buffers per LAB chip).
When a Level~3 trigger is issued, sampling will be frozen for the 8 blocks of SCAs to digitize data, while the remaining 24 blocks continue to sample.

\section{Acknowledgments}

We are grateful to NASA for their support for ANITA through 
Grant NNX15AC20G. We thank the U.S. National 
Science 
Foundation-Office of
Polar Programs. A. Connolly would like to thank the National Science 
Foundation for their support through CAREER
award 1255557. This work was also supported by 
collaborative visits 
funded by the Cosmology and Astroparticle Student and 
Postdoc Exchange Network (CASPEN).
Lastly, we thank Brian Clark 
and Suren Gourapura for their valuable feedback.

\section{References}

\bibliography{tuff_nim} 

\begin{thebibliography}{7}
\providecommand{\natexlab}[1]{#1}
\providecommand{\url}[1]{\texttt{#1}}
\providecommand{\urlprefix}{URL }
\expandafter\ifx\csname urlstyle\endcsname\relax
  \providecommand{\doi}[1]{doi:\discretionary{}{}{}#1}\else
  \providecommand{\doi}[1]{doi:\discretionary{}{}{}\begingroup
  \urlstyle{rm}\url{#1}\endgroup}\fi
\providecommand{\bibinfo}[2]{#2}

\bibitem[{Gorham et~al.(2009)}]{instrPaper}
\bibinfo{author}{P.~W. Gorham}, et~al., \bibinfo{title}{{The Antarctic
  Impulsive Transient Antenna Ultra-high Energy Neutrino Detector Design,
  Performance, and Sensitivity for 2006-2007 Balloon Flight}},
  \bibinfo{journal}{Astropart. Phys.} \bibinfo{volume}{32}
  (\bibinfo{year}{2009}) \bibinfo{pages}{10--41},
  \doi{\bibinfo{doi}{10.1016/j.astropartphys.2009.05.003}}.

\bibitem[{Askar'yan(1962)}]{askaryan}
\bibinfo{author}{G.~A. Askar'yan}, \bibinfo{title}{{Excess negative charge of
  an electron-photon shower and its coherent radio emission}},
  \bibinfo{journal}{Sov. Phys. JETP} \bibinfo{volume}{14}~(\bibinfo{number}{2})
  (\bibinfo{year}{1962}) \bibinfo{pages}{441--443}, \bibinfo{note}{[Zh. Eksp.
  Teor. Fiz.41,616(1961)]}.

\bibitem[{Gorham et~al.(2007)}]{askaryan_observation}
\bibinfo{author}{P.~W. Gorham}, et~al., \bibinfo{title}{{Observations of the
  Askaryan effect in ice}}, \bibinfo{journal}{Phys. Rev. Lett.}
  \bibinfo{volume}{99} (\bibinfo{year}{2007}) \bibinfo{pages}{171101},
  \doi{\bibinfo{doi}{10.1103/PhysRevLett.99.171101}}.

\bibitem[{Matassa(2011)}]{milsat}
\bibinfo{author}{C.~K. Matassa}, \bibinfo{title}{Comparing the capabilities and
  performance of the ultra high frequency follow-on system with the mobile user
  objective system}, Master's thesis, \bibinfo{school}{Naval Postgraduate
  School}, \urlprefix\url{https://calhoun.nps.edu/handle/10945/5711},
  \bibinfo{year}{2011}.

\bibitem[{Varner et~al.(2007)Varner, Ruckman, Gorham, Nam, Nichol, Cao, and
  Wilcox}]{labrador}
\bibinfo{author}{G.~S. Varner}, \bibinfo{author}{L.~L. Ruckman},
  \bibinfo{author}{P.~W. Gorham}, \bibinfo{author}{J.~W. Nam},
  \bibinfo{author}{R.~J. Nichol}, \bibinfo{author}{J.~Cao},
  \bibinfo{author}{M.~Wilcox}, \bibinfo{title}{{The large analog bandwidth
  recorder and digitizer with ordered readout (LABRADOR) ASIC}},
  \bibinfo{journal}{Nucl. Instrum. Meth.} \bibinfo{volume}{A583}
  (\bibinfo{year}{2007}) \bibinfo{pages}{447--460},
  \doi{\bibinfo{doi}{10.1016/j.nima.2007.09.013}}.

\bibitem[{{Nishimura} et~al.(????){Nishimura}, {Andrew}, {Cao}, {Cooney},
  {Gorham}, {Macchiarulo}, {Ritter}, {Romero-Wolf}, and {Varner}}]{ritc}
\bibinfo{author}{K.~{Nishimura}}, \bibinfo{author}{M.~{Andrew}},
  \bibinfo{author}{Z.~{Cao}}, \bibinfo{author}{M.~{Cooney}},
  \bibinfo{author}{P.~{Gorham}}, \bibinfo{author}{L.~{Macchiarulo}},
  \bibinfo{author}{L.~{Ritter}}, \bibinfo{author}{A.~{Romero-Wolf}},
  \bibinfo{author}{G.~{Varner}}, \bibinfo{title}{{A low-resolution, GSa/s
  streaming digitizer for a correlation-based trigger system}},
  \bibinfo{journal}{ArXiv e-prints 1203.4178} .

\bibitem[{Roberts et~al.(2017)Roberts, Oberla, Allison, Varner, Spack, Fox, and
  Rotter}]{lab4d}
\bibinfo{author}{J.~Roberts}, \bibinfo{author}{E.~Oberla},
  \bibinfo{author}{P.~Allison}, \bibinfo{author}{G.~Varner},
  \bibinfo{author}{S.~Spack}, \bibinfo{author}{B.~Fox},
  \bibinfo{author}{B.~Rotter}, \bibinfo{title}{{LAB4D: A Low Power,
  Multi-GSa/s, Transient Digitizer with Sampling Timebase Trimming
  Capabilities}}, \bibinfo{journal}{Nucl. Instrum. Meth.} \bibinfo{volume}{1}
  (\bibinfo{year}{2017}) \bibinfo{pages}{12}.

\end{thebibliography}
\bibliographystyle{elsarticle-num-names.bst}

\end{document}